\documentclass[preprint]{aa}
\usepackage{graphicx}
\usepackage{aas_macros}
\usepackage{natbib}
\usepackage{amsmath}
\usepackage{amssymb}
\usepackage{setspace}
\usepackage[utf8x]{inputenc}
\usepackage{latexsym}
\usepackage{color}
\usepackage{dcolumn}
\usepackage{bm}

\bibpunct{(}{)}{;}{a}{}{,}

\begin{document}
 
\title{Searching for the first stars with the Gaia mission} 

\author{R. S. de Souza\inst{1,2,3} 
	\and A. Krone-Martins\inst{4}     \and  E.E.O. Ishida\inst{1,3} \and B. Ciardi\inst{3}}
\offprints{Rafael S. de Souza \email{rafael@kasi.re.kr}}

\institute{$^{1}$IAG, Universidade de S\~{a}o Paulo, Rua do Mat\~{a}o 1226,  05508-900, S\~{a}o Paulo, SP, Brazil\\
$^{2}$Korea Astronomy \& Space Science Institute, 305-348, Daejeon, Korea\\
$^{3}$Max-Planck-Institut f\"ur Astrophysik, Karl-Schwarzschild-Str. 1, D-85748 Garching, Germany\\
$^{4}$SIM, Faculdade de Ci\^encias, Universidade de Lisboa, Ed. C8, Campo Grande, 1749-016, Lisboa, Portugal\\
}

\date{Accepted -- Received}

\date{Released Xxxxx XX}

\authorrunning{de Souza \textit{et al.}}
   \titlerunning{ Pop III Orphan Afterglows}

\abstract
{}
{We construct a  theoretical model to predict the number of orphan afterglows (OA) from  gamma-ray bursts (GRBs)
triggered by  primordial  metal-free  (Pop III) stars  expected to be observed by the Gaia mission. In particular, we consider primordial metal-free stars  that were affected by radiation from other stars (Pop III.2) as a possible target.
}
{We use a  semi-analytical approach that includes  all relevant feedback effects   to construct  cosmic  star formation history and its connection with  the cumulative number of GRBs. The OA events are generated using the  Monte Carlo method, and realistic simulations of Gaia's scanning law are performed  to derive the  observation probability expectation.} 
{We show that Gaia can observe  up to 2.28 $\pm$ 0.88  off-axis afterglows and  2.78 $\pm$ 1.41  on-axis during the five-year nominal mission. This implies that a  nonnegligible percentage of afterglows   that may be  observed by  Gaia ($\sim 10\%$) could  have  Pop III stars as progenitors.  }
{}

\maketitle
\section{Introduction}
The first stars (hereafter, Pop~III-primordial metal-free) in the Universe are thought to have 
played a crucial role in the early cosmic evolution 
by emitting the first light and producing the first heavy elements
\citep{bromm09}.  Understanding  such objects is very important since their 
detection would permit  the pristine regions of the Universe to be probed. 
However, there has been no direct observation of the so-called
Pop~III  stars up to now.

Pop III stars may produce 
collapsar gamma-ray bursts (GRBs) whose total isotropic energy 
could be $\approx 2$ orders of magnitude larger than average
\citep{barkov2010,komissarov2010,meszaros2010,suwa2011,toma2011}. 
Even if the Pop III star has a supergiant
hydrogen envelope, 
the GRB jet can break out of it because of the long-lasting 
accretion of the envelope itself \citep{nagakura2011,suwa2011}. 
It is of great importance to study the rate and 
detectability of Pop~III GRB prompt emissions and afterglows in current and future surveys.  We explore here the  possibility to observe these objects through  their  afterglows \citep{toma2011}.   
 Observations of GRB afterglows  
make it possible to derive physical properties of the explosion mechanism
and the circumburst medium. It is intriguing to search for signatures of metal-poor stars
 in the GRB afterglows at low and high redshifts.

GRB optical afterglows are one of the possible transients to be detected
by the Gaia\footnote{http://www.rssd.esa.int/GAIA/} mission. Recently  \citet{Japelj2011}  explored the detectability of such afterglows with Gaia using a Monte Carlo approach that inspired us.  As the GRB jet sweeps the interstellar medium, the Lorentz factor of the jet is
decelerated and the  jet starts to expand sideways, eventually  becoming detectable by off-axis observers. 
These afterglows are not associated with the prompt GRB emission and are called orphan afterglows (OA) \citep{nakar2002, rossi2008}.

De Souza et al. (2011) showed that, considering EXIST\footnote{http://exist.gsfc.nasa.gov/design/} specifications, we can expect to observe a maximum of $\approx 0.08$ GRBs with $z>10$  per year originating from primordial metal-free stars (Pop III.1) and $\approx 20$ GRBs with $z>6$ per year coming from primordial metal-free stars that were affected by the radiation from other stars (Pop III.2). In the context of the current \textit{Swift}\footnote{http://swift.gsfc.nasa.gov/docs/swift/swiftsc.html} satellite, $\approx 0.2$ GRBs with $z>6$ per year from Pop III.2 stars are expected.  These numbers reflect the fact that, compared to Pop~III.1 stars, Pop III.2 stars  are more abundant and can be observed in a lower redshift range, which makes them more suitable targets. In  the light of such results, the calculations presented here will focus on Pop III.2 stars alone. 

Searches have been made of OAs  by both X-ray surveys \citep{grindlay1999,greiner2000} and  optical searches \citep{becker2004,rykoff2005,rau2006,malacrino2007}.  
The purpose of the present paper is to calculate the Pop~III.2 GRB OA  rate  
that might be detected by the Gaia mission \citep[for  more details about Gaia,  see,  e.g.,][]{Perryman:2001p3838, Lindegren:2009p8828}. 

The Gaia mission is one of the most ambitious projects of modern astronomy. It aims to  create  a very precise tridimensional, dynamical, and chemical  census of our Galaxy from astrometric, spectrophotometric,  and spectroscopic data. In order to do this, the Gaia satellite will perform observations of the entire sky in a continuous scanning created from the coupling of rotations and precession movements called  the scanning law. For point sources, these observations will be unbiased and the data of all the objects bellow a certain limiting magnitude (G=20)  will be transferred to the ground. Certainly, galactic and extragalactic sources will be among those objects. 

Typically, Pop~III.2 stars are formed
in an initially ionized gas \citep{Johnson06,Yoshida07}. 
They are thought to be 
less massive  than Pop~III.1 stars
but still massive enough to produce GRBs.
Recent results from \citet{greif2011} show that,  
instead of forming a single object, the gas in mini-halos fragments vigorously into a number of protostars with a range of different  masses. It is not clear up to now how this initial range of mass will be  mapped into the final mass function of Pop III stars.  The most likely conclusion is that Pop III stars are less likely to reach masses in excess of $\sim 140 M_{\odot}$, which consequently affect the estimated number of GRBs from Pop III.1.  \citet{hosokawa2011}, performing state-of-the-art radiation-hydrodynamics simulations, 
showed that the typical mass of Pop III stars  could be $\sim 43 M_{\odot}$.
 Here we assume that this will not affect significantly the mass range  assumed for Pop III.2 ($\sim 10-100 M_{\odot}$).

The paper is organized as follows.
In Sect. 2, we  calculate 
the formation rate of primordial GRBs.  
In Sect. 3, we calculate the OA light curves and their redshift distribution.  
In Sect. 4, we discuss the details of the   Gaia mission and  derive the probability of a given event to be observed by Gaia. 
In Sect. 5,  we discuss the results,  and  finally,  in Sect. 6,  we give our concluding remarks.
Throughout the paper,  we adopt the standard $\Lambda$ cold dark matter 
model with the best-fit cosmological parameters 
from   \citet{jarosik2010} 
(WMAP-Yr7\footnote{http://lambda.gsfc.nasa.gov/product/map/current/}),  
$\Omega_{\rm m} = 0.267, \Omega_{\Lambda} = 0.734$,  
and $H_0 = 71 {\rm km}~{\rm s}^{-1}{\rm Mpc}^{-1}$.

\section{GRB redshift distribution}

 To estimate the formation rate of GRBs from Pop III stars at a given redshift,  we closely follow \citet{rafael2011}. 
 Since long GRBs  are expected to follow the death of  very massive  stars,  their rate could provide a useful probe for  cosmic star formation history  (SFH)  \citep[e.g.,][]{totani1997,ciardi2000,bromm2002,conselice2005,campisi2010, campisi2011,ishida2011, rafael2011,robertson2011}. 
 However, the connection between the star formation rate (SFR)  density  and GRB rate is not clearly understood and can be redshift dependent \citep[e.g.,][]{yuksel2008,kistler2009,robertson2011}.   Since host galaxies of long-duration GRBs are often observed to be metal poor. 
Several studies connect the origin of long GRBs with the metallicity of their progenitors \citep[e.g.,][]{meszaros2006,woosley2006,Salvaterra2007,Salvaterra2009,campisi2011b}. Consequently,  the GRB-SFR connection  could be dependent on the cosmic metallicity evolution. However,  this connection is not   yet completely understood, since there is also evidence  of regions within GRB host galaxies known to possess higher metallicities \citep{levesque2010}. 
 
  Despite  such uncertainties,  we expect the connection between SFR and GRBs to be less affected by this effect because Pop III stars and their environment are metal poor. In other words, Pop III stars are more likely to produce GRBs than ordinary stars. It is important to keep in mind that any  prediction will be convolved with systematic effects  that we are not taking into account. However, as pointed out in \citet{ishida2011}, the assumption is good enough to agree with available  observational data.

We implicitly  assume that the formation rate of long GRBs (duration longer than 2 sec) follows closely the SFH. 
The number of GRBs per comoving volume per time can be expressed as
\begin{equation}
 \Psi_{\rm GRB}(z) = \eta_{\rm GRB}\Psi_{*}(z),
 \label{psigrbreal}
 \end{equation}
where $\eta_{\rm GRB}$ is the GRB formation efficiency  and  $\Psi_{*}$ is the SFR.  
 Over a particular time interval,  $\Delta t_{\rm obs}$,  in the observer rest frame, 
the number of GRBs originating between redshifts $z$ and $z + dz$
is
\begin{equation}
\frac{{\rm d}N_{\rm GRB}}{{\rm d}z} = \Psi_{\rm GRB}(z)\frac{\Delta t_{\rm obs}}{1+z}
\frac{{\rm d}V}{{\rm d}z},
\label{dngrbtrue}
\end{equation}
where ${\rm d}V/{\rm d}z$ is the comoving volume element per  redshift unit.

 \subsection{Star Formation History}
 
To estimate the SFR at early epochs, we assume  that stars are formed in collapsed dark matter halos  \citep[for more details, please see][]{rafael2011}. 
The number of collapsed
objects is given by  the halo mass function 
\citep{Hernquist2003,greif2006,trenti2009}.  In what follows, we adopt the  Sheth-Tormen function, $f_{\rm ST}$ \citep{sheth1999}. 
To estimate the fraction of mass inside each halo  capable of collapsing  and forming stars,  we 
include  the following important feedback mechanisms:

\begin{enumerate}

\item $\rm H_2$ Photodissociation

Hydrogen molecules (H$_{2}$) are the primary coolant
in the gas within  small-mass ``mini-halos." 
H$_{2}$ are also sensitive  to ultra-violet radiation in the Lyman-Werner (LW) bands 
and can easily be suppressed  by it. 
We model the dissociation effect by setting the minimum mass for halos that 
are able to host Pop~III stars \citep{yoshida2003}.

\item Reionization

 Inside growing {H\sc{ii}}  regions, the gas is highly ionized and 
the temperature is $\sim 10^4$ K. The volume-filling factor of ionized regions, $Q_{\rm {H\sc{II}}}(z)$,  determines when  the formation of Pop III.1 stars 
is terminated and switches 
to Pop III.2. To calculate $Q_{\rm {H\sc{II}}}(z)$, we closely follow \citet{wyithe2003} as in \citet{rafael2011}.

\item Metal Enrichment 

Metal enrichment in the intergalactic medium (IGM)
determines when the formation of primordial  
stars is terminated (locally) and 
switches from the Pop~III mode to a more conventional mode of star formation.  
We assume that star-forming halos launch a wind of metal-enriched
gas at 
$z\gtrsim20$.  Then we  follow the metal-enriched wind propagation outward from a central galaxy with a  given velocity $v_{wind}$, traveling over a comoving distance $R_{wind}$. We estimate the ratio of gas mass enriched by the wind to the total gas mass in each halo, and then we evaluate the average metallicity over cosmic scales as a function of redshift. 
We effectively assume that the so-called critical 
metallicity is very low \citep{schneider2002,schneider2003,bromm2003,omukai2005,frebel2007,Krzysztof2010}. 
Therefore,   Pop~III stars
are not formed in a metal-enriched region, regardless of the 
actual metallicity.

\end{enumerate}
 
\citet{rollinde2009MNRAS} investigated the role of Pop III stars in the cosmic metallicity  evolution, in particular, the local metallicity function of  the Galactic halo. They show that Pop III SFR should not be larger than  $3\times 10^{-3} M_{\odot} yr^{-1} Mpc^{-3}$ at any redshift.    
We also include this additional constraint as an upper limit for our optimistic model. 

\begin{figure}
\includegraphics[width=0.9\columnwidth]{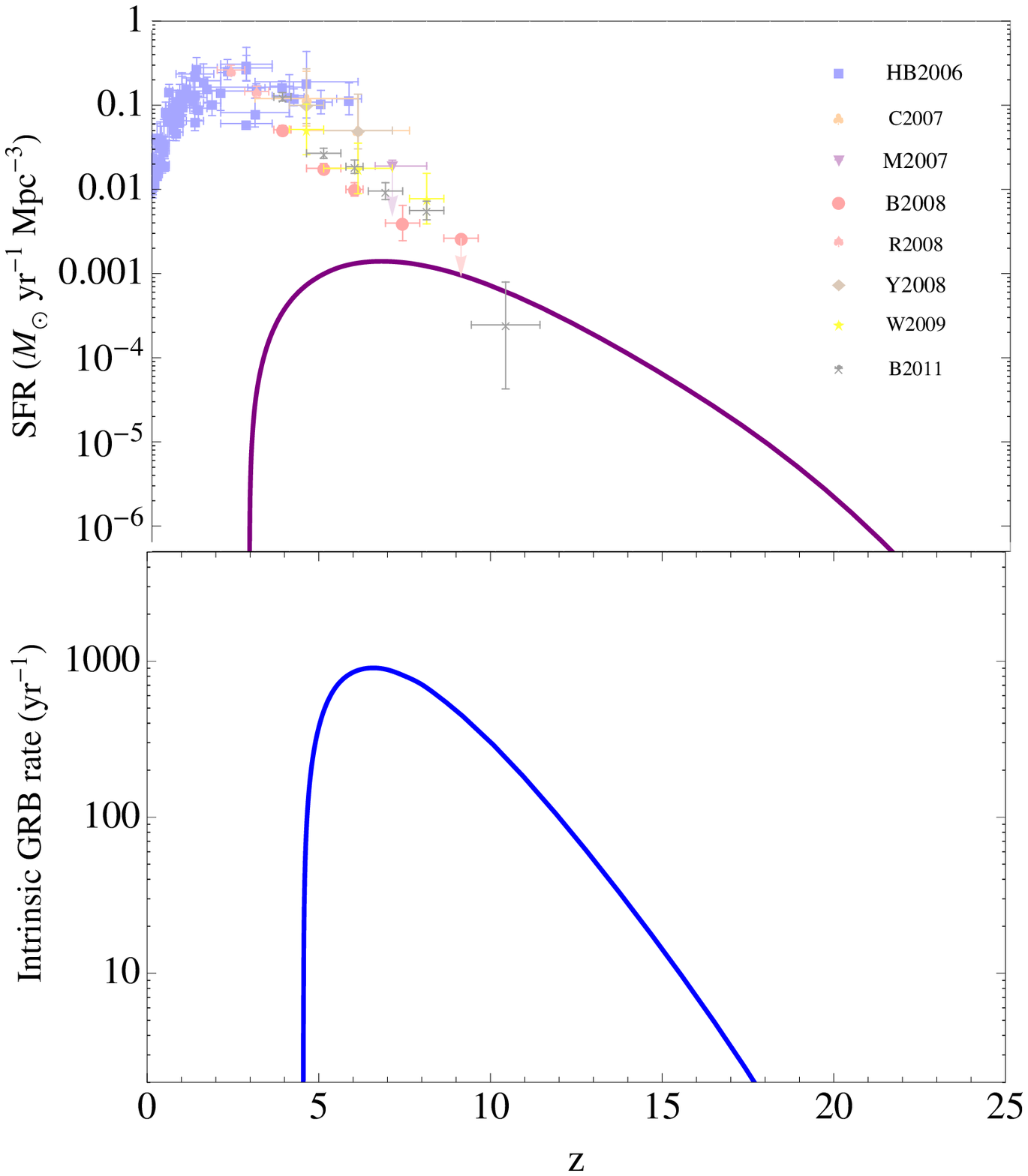}
\caption
{Top: optimistic model for Pop~III.2 star formation rate (SFR)  assuming  a  high star formation efficiency  and
low chemical enrichment. The light points are  independent SFR determinations compiled from the literature. }
{Bottom:  intrinsic GRB rate  ${\rm d}N_{\rm GRB}/{\rm d}z$, i.e.,  the number of  GRBs per year on the sky
 (on-axis + off-axis) according to Eq. (\ref{dngrbtrue}). 
This represents our optimistic model assuming a high star formation 
efficiency for Pop III.2, slow chemical enrichment,  GRB formation efficiency of $f_{GRB}= 0.001$ and a Salpeter IMF.}  
\label{fig:SFRII}
\end{figure}

The top panel of Fig.  \ref{fig:SFRII} shows  the  upper limit  for   Pop~III.2  SFR,   based on  \citet{rafael2011} with the additional  constraints cited above.  
 The Pop~III.2 SFR is compared  with a compilation of independent measures from \citet{hopkins2006} up to $z \approx 6$ 
and from observations of color-selected Lyman Break Galaxies \citep{mannucci2007, bouwens2008, bouwens2011},  UV+IR measurements \citep{reddy2008},  and GRB observations \citep{chary2007, yuksel2008, wang2009} at higher $z$ (in the figure, these will be refereed to as H2006, M2007, B2008, B2011, R2008, C2007, Y2008,  and W2009, respectively).

\subsection{Initial Mass Function and GRB Formation Efficiency}

The stellar initial mass function (IMF) is critically
important to determine the Pop~III GRB rate.  The IMF  determines the fraction of stars with minimum mass that are  able to trigger 
GRBs,  $\sim 25 M_{\odot}$ \citep{bromm2006}. The  $f_{\rm GRB}$  factor gives the fraction of stars in this range of mass that will produce GRBs. 

The GRB formation efficiency factor per stellar mass is 
\begin{equation}
\eta_{\rm GRB} =  f_{\rm GRB} \frac{\int_{M_{\rm GRB}}^{M_{\rm up}}\phi(m)dm}
{\int_{M_{\rm low}}^{M_{\rm up}}m\phi(m)dm}, 
\label{etagrb}
\end{equation}
where $\phi(m)$ is the stellar IMF for which we considered a  power law with the standard Salpeter slope
$\phi (m) \propto m^{-2.35},$
$M_{\rm low}$ and $M_{\rm up}$ are the minimum and maximum mass 
for a given stellar type (respectively $10 M_{\odot}$ and $\sim 100 M_{\odot}$ for Pop~III.2), and
$M_{\rm GRB}$ is the minimum mass  able to trigger 
GRBs, which we set to be $25 M_{\odot}$ \citep{bromm2006}. 

De Souza et al. (2011) placed upper limits on the intrinsic GRB rate (including the off-axis GRB). In the following, we   
set $f_{GRB} = 0.001$ and  $\eta_{\rm GRB}/f_{\rm GRB} \sim 
1/87 M_{\odot}^{-1}$ as an optimistic case, consistent with their results.    
The bottom panel of Fig. \ref{fig:SFRII} shows the optimistic case  for intrinsic GRB rate  adopted in this work.

\section{Number of Observed Orphans}

\subsection{Afterglow Model}

To calculate the  afterglow light curves of Pop~III GRBs,   we 
follow  the standard 
prescription from  \citet{sari1998,sari1999} and \citet{meszaros2006}. 
The spectrum consists of power-law segments linked by critical 
break frequencies. 
These are $\nu_{\rm a}$ (the self-absorption frequency), 
$\nu_{\rm m}$ (the peak of injection frequency),  
and $\nu_{\rm c}$ (the cooling frequency),  given by
\begin{eqnarray}
\nu_{\rm m} &\propto& ~ (1+z)^{1/2} g(p)^2
\epsilon_e^2 \epsilon_B^{1/2} E_{\rm iso}^{1/2} t_d^{-3/2}, \label{eq:numt}\nonumber\\
\nu_{\rm c} &\propto& ~ (1+z)^{-1/2} \epsilon_B^{-3/2}
n^{-1} E_{\rm iso}^{-1/2} t_d^{-1/2}, \label{eq:nuct}\nonumber\\
\nu_{\rm a} &\propto& ~ (1+z)^{-1} \epsilon_e^{-1}
\epsilon_B^{1/5} n^{3/5} E_{\rm iso}^{1/5}, \label{eq:nuat}\nonumber\\
F_{\nu,{\rm max}} &\propto& ~ (1+z) \epsilon_B^{1/2} n^{1/2} E_{\rm iso}
d_{L}^{-2},~ \label{eq:Fnumaxt}
\end{eqnarray}
where $g(p) = (p-2)/(p-1)$  is a function   of the energy spectrum index 
of electrons $(N(\gamma_e)d\gamma_e\propto \gamma_e^{-p}d\gamma_e$, where $\gamma_{e}$ is the electron Lorentz factor),     
$\epsilon_{e}$ and $\epsilon_B$ are the efficiency factors \citep{meszaros2006}, $E_{iso}$ is the isotropic kinetic  energy, n is the density of the medium,  and $F_{\nu,{\rm max}}$ is the observed peak flux at  luminosity distance $d_{L}$ from the source. 

There are two types of spectra. 
If $\nu_{\rm m} < \nu_{\rm c}$, we call it the
{\it slow cooling case}. The flux at the observer, $F_\nu$, is given by
\begin{equation}
\label{spectrumslow}
F_\nu=\left\{ \begin{array}{ll}(\nu_{\rm a} / \nu_{\rm m} )^{1/3}(\nu/\nu_{\rm a})^2 F_{\nu,{\rm max}}, &
\nu_{\rm a}>\nu, \\ 
( \nu / \nu_{\rm m})^{1/3} F_{\nu,{\rm max}}, &
\nu_{\rm m}>\nu>\nu_{\rm a}, \\
( \nu / \nu_{\rm m} )^{-(p-1)/2} F_{\nu,{\rm max}}, &
\nu_{\rm c}>\nu>\nu_{\rm m}, \\ 
( \nu_{\rm c} / \nu_{\rm m} )^{-(p-1)/2} ( \nu / \nu_{\rm c})^{-p/2}
F_{\nu,{\rm max}}, & \nu>\nu_{\rm c}. \end{array}\right.
\end{equation}

For $\nu_{\rm m}>\nu_{\rm c}$, called the \textit{fast cooling case}, the spectrum is

\begin{equation}
\label{spectrumfast}
F_\nu=\left\{ \begin{array}{ll}(\nu_{\rm a} / \nu_{\rm c} )^{1/3}(\nu/\nu_{\rm a})^2 F_{\nu,{\rm max}}, &
\nu_{\rm a}>\nu, \\ 
( \nu / \nu_{\rm c})^{1/3} F_{\nu,{\rm max}}, &
\nu_{\rm c}>\nu>\nu_{\rm a}, \\
( \nu / \nu_{\rm c} )^{-1/2} F_{\nu,{\rm max}}, &
\nu_{\rm m}>\nu>\nu_{\rm c}, \\ 
( \nu_{\rm m} / \nu_{\rm c} )^{-1/2} ( \nu / \nu_{\rm m})^{-p/2}
F_{\nu,{\rm max}}, & \nu>\nu_{\rm m}. \end{array}\right.
\end{equation}

 Initially the jet propagates as if it were spherical with
an equivalent isotropic energy of $E_{\rm true} = \theta_{j}^2E_{\rm iso}/2$, where $\theta_{j}$ is the half-opening angle of the  jet. 
Even if the prompt emission is  highly collimated,  
the Lorentz factor drops $\gamma_{\rm d} < \theta_{j}^{-1}$ around the time
\begin{equation}
t_{\theta} \sim 2.14  \left(\frac{E_{\rm iso}}{5\times 10^{54}}\right)^{1/3}\left(\frac{\theta_{j}}{0.1}\right)^{8/3}n^{-1/3}(1+z)~ \rm days, 
\end{equation}
and  the jet starts to expand sideways \citep{ioka2005}.  Consequently, the jet  becomes detectable by the off-axis observers. 
These afterglows are not associated with the prompt GRB emission.

 Due to relativistic beaming, an
observer  located at $\theta_{obs}$, outside the initial opening
angle of the jet ($\theta_{obs}> \theta_j$), will observe the afterglow emission only at $t \sim t_\theta$, 
 when $\gamma_{\rm d} = \theta_{j}^{-1}$.
 
\begin{figure}
\includegraphics[width=1\columnwidth]{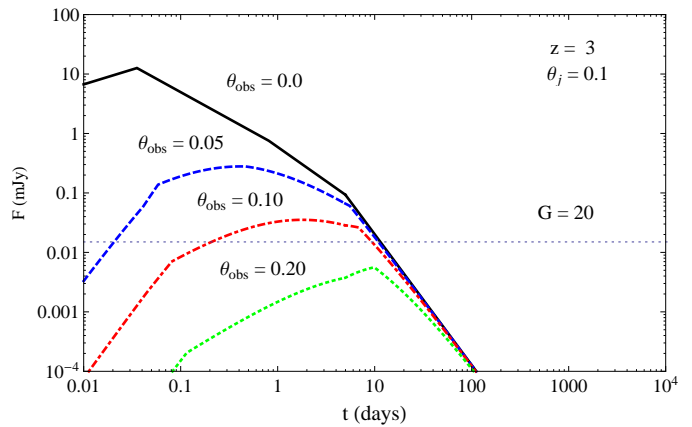}
\caption{Example of afterglow light curve at $z=3$ as a function of observed angle, $\theta_{obs}$. We show the evolution of afterglow flux $F (mJy)$ as a function of time $t$ (days) and observed angle $\theta_{obs}$ for typical parameters: isotropic kinetic energy $E_{iso} = 10^{54}$ erg, electron spectral index $p = 2.5$, plasma parameters $\epsilon_{e} = 0.1$, $\epsilon_{B} = 0.01$, half-opening angle jet $\theta_{j} = 0.1$,  interstellar medium density $n = 1 cm^{-3}$ and frequency $\nu = 5\times 10^{14} \rm{Hz}$.  The horizontal dotted line is the integrated Gaia flux limit; solid black line, $\theta_{obs} = 0$;  dashed blue line, $\theta_{obs} = 0.05$; dot-dashed red line, $\theta_{obs} = 0.1$; dotted green line, $\theta_{obs} = 0.20$.}
\label{fig:GRBafterglow}
\end{figure}

The received afterglow flux by an off-axis observer in the point source approximation, valid for  $\theta_{obs} \gg \theta_{j}$,  is related to that seen by an on-axis observer by \citep{granot2002,totani2002,Japelj2011}
\begin{equation}
F_{\nu}(\theta_{obs},t) = \xi^3F_{\nu/\xi}(0,\xi t),
\end{equation}
where
\begin{equation}
\xi \equiv (1-\beta)/(1-\beta\cos\theta_{obs}), 
\end{equation}
and $\beta = \sqrt{1-1/\gamma_{\rm d}^2}$.  The time evolution of the Lorentz factor in given by

\begin{equation}
\label{gamma}
\gamma_{\rm d}(t) = \left \lbrace
\begin{array}{ll}
\theta_{\rm j}^{-1}\left( \frac{t}{t_{\rm j}}\right)^{-3/8} & t < t_{\rm j}\\
\theta_{\rm j}^{-1}\left( \frac{t}{t_{\rm j}}\right)^{-1/2} & t > t_{\rm j},\\
\end{array}
\right.
\end{equation}
where $t_j$ is the jet break time, $\approx 0.7(1+z)(E_{51}/n)^{1/3}(\theta_j/0.1)^2$ days \citep{sari1999}. Figure \ref{fig:GRBafterglow} shows four  examples of  afterglows as a function of observed angle $\theta_{obs}$ for the case of $\theta_{j} = 0.1$ at $z =3$ for typical parameters described in the figure.  The flux is calculated for an observational frequency $\nu = 5\times 10^{14} \rm{Hz}$ within the Gaia bandwidth. Depending on the parameters of the afterglow, the light curve can appear above the Gaia observational limits. Due to the large quantity of free parameters, a Monte Carlo approach is essential to explore the detectability of a large number of events and will be explained in the next section.

\subsection{Dust Extinction}

A fraction of GRBs with X-ray or radio afterglows can be hidden by  dust absorption from their host galaxies. 
The observed flux after extinction correction can be simply written as \citep[see, e.g,][]{eliasdottir2009}
\begin{equation}
F_{\nu}^{dust} = F_{\nu}(\theta_{obs},t)10^{-0.4A_{\lambda}},
\end{equation}
where $A_{\lambda}$ is the extragalactic extinction  along the line of sight, $A_{\lambda;ext}$,   as a function of the wavelength $\lambda$ plus the extinction from the Milk Way, $A_{\lambda;MW}$. 

\begin{figure}
\includegraphics[width=1\columnwidth]{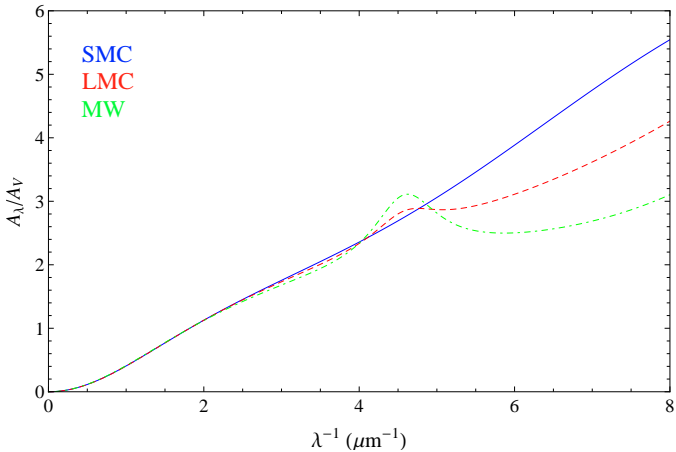}
\caption{Extinction laws usually adopted in literature: the Small  Magellanic Cloud (SMC) law (blue line), the Large  Magellanic Cloud (LMC) law (red dashed line) and the Milky Way  (MW) law (green dot-dashed line).}
\label{fig:dust}
\end{figure}

\begin{figure}
\includegraphics[width=1\columnwidth]{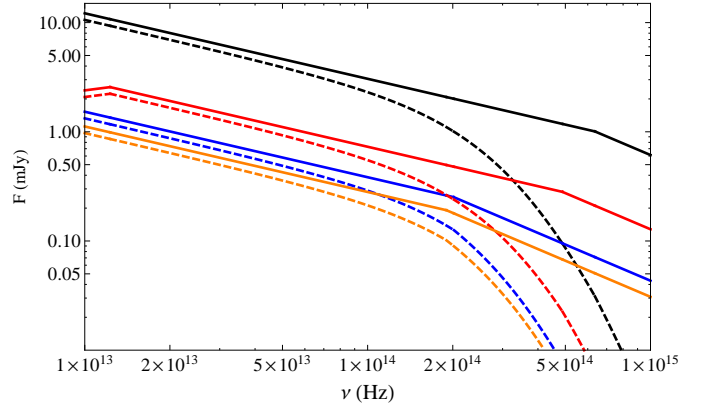}
\caption{Example of a spectral energy distribution for observed GRBs with (dashed lines) and without (solid lines) extinction, assuming the SMC extinction law. We show the  afterglow flux $F (mJy)$ as a function of frequency $\nu$ (Hz) for typical parameters: isotropic kinetic energy $E_{iso} = 10^{54}$ erg, electron spectral index $p = 2.5$, plasma parameters $\epsilon_{e} = 0.1$, $\epsilon_{B} = 0.01$, half-opening angle jet $\theta_{j} = 0.1$,  and interstellar medium density $n = 1 cm^{-3}$.   The black line represents $\theta_{obs} = 0$, t = 0.5 days,  blue line, $\theta_{obs} = 0$, t = 5 days, red line, $\theta_{obs} = 0.05$, t = 0.5 days,  and orange line, $\theta_{obs} = 0.05$, t = 5 days.}
\label{fig:SED}
\end{figure}

For $A_{\lambda;ext}$, we adopted a simple Small  Magellanic Cloud (SMC) type extinction model. The SMC  model was already shown to provide good fits for several GRB afterglows observations \citep[see, e.g,][]{eliasdottir2009}. For $A_{\lambda;MW}$, we use the average value 0.15 from observations of \citet{schady2012} and adopt a typical value of  0.3  for $A_{V}$.   In Fig. \ref{fig:dust}, we show the SMC extinction curve in comparison with other popular models,  Large  Magellanic Cloud (LMC) and Milky Way  (MW). The model choice has no significant effect on our results, since all of them have a similar trend  in the G band range. In Fig. \ref{fig:SED},  we show the effect of dust extinction in the spectral energy distribution (SED) of GRBs. The effect is significant in the G band ($\sim 5 \times 10^{14}$ Hz), which will considerably decrease the detection rate of optical GRBs,  mainly at high-z.

\subsection{IGM and DLA absorption}

For high-z GRBs, much of the optical and near-infrared light will be absorbed  by the $Ly\alpha$ forest, which provides a powerful tool to probe the reionization era \citep{Miralda1998,Mesinger2008,Ciardi2011}. The GRB 050904 at z = 6.3 was  the first  attempt to probe the IGM through GRBs at the epoch of reionization by using  the damping wing at wavelengths redward of the Lyman break \citep{Totani2006,Kawai2006}. 
The absorption by the neutral IGM can be approximated by \citep{McQuinn2008}  
\begin{eqnarray}
\tau_{IGM} \approx 900~ kms^{-1}~x_{\rm HI}\left(\frac{1+z_{host}}{8}\right)^{3/2}\times \\
\left(\frac{H(z_{host})R_b}{(1+z_{host})}-c\frac{\nu_z-\nu_{\alpha}}{\nu_{\alpha}}\right)^{-1},\nonumber
\end{eqnarray}
where  $\nu_{\alpha}$ is the rest frame of the $\rm Ly{\alpha}$ line,  $R_b$ represents the size of an HII region surrounded by an IGM with neutral fraction $x_{\rm HI}$,  $z_{host}$ is the redshift of the GRB host galaxy, and H(z) is the Hubble parameter for  a $\Lambda$CDM cosmology. To estimate $x_{\rm HI}$,   we use the prescription detailed in \citet{rafael2011} (see Fig.1). 
The optical depth of the damped Ly$\alpha$ absorber (DLA),  $\tau_{DLA}=N_{\rm HI} \sigma_{\alpha}[\nu_{obs}(1+z_{host})]$,  can be computed by
\begin{eqnarray}
\tau_{DLA} = 7.26\left(\frac{N_{\rm HI}}{{10^{21} cm^{-2}}}\right)\left(\frac{1+z_{obs}}{8}\right)^4\times\\
\left(\frac{1+z_{host}}{8}\right)^{-2}\left(\frac{\Delta\lambda}{20\rm \AA\ }\right)^{-2},\nonumber 
\end{eqnarray}
 (e.g.,  \citealt{Barkana2004}) where $\nu_{obs} = c/\lambda_{obs}$,  $(1+z_{obs}) = \lambda_{obs}/\lambda_{\alpha}$,  $N_{\rm HI}$ is the total column density of $\rm HI$ in the host galaxy, and $\lambda_{obs}= \Delta\lambda+\lambda_{\alpha}(1+z)$. $N_{\rm HI}$ is randomly chosen assuming a cumulative distribution function  scaling  as $N_{\rm HI}^{0.3}$,  between $10^{18}$ and 
$10^{21.5} cm^{-2}$ (see \citealt{Chen2007,McQuinn2008}). For each event, 
 $R_b$ is  chosen  from a lognormal distribution  between 1-100 Mpc motivated by a visual  inspection in Fig. 5 from \citet{McQuinn2008}. 

\subsection{Mock sample} 
The mock sample is generated by the Monte Carlo method assuming different probability distribution functions (PDF)  for each quantity as explained below.  
The medium density $n$ is randomly chosen from a flat distribution within $0.1-1 cm^{-3}$.

\subsubsection{Redshift PDF}

\begin{figure}
\includegraphics[width=1\linewidth]{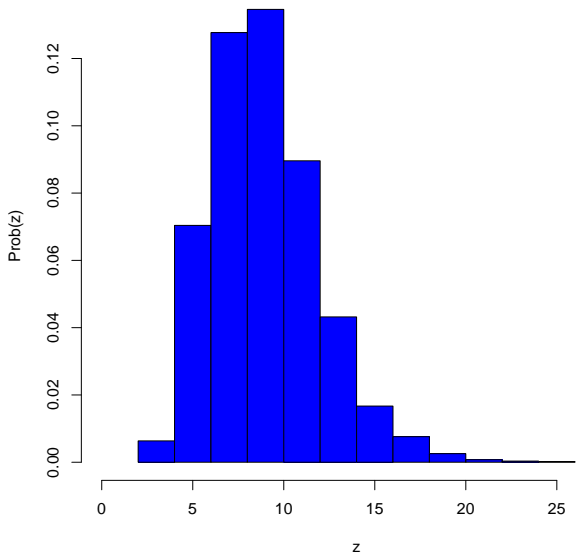}
\caption{Redshift PDF. Probability of a given event  appearing in a certain range of redshift. }
\label{fig:probz}
\end{figure}

We generate the GRB events  randomly  in redshift with a PDF given by  Eq. (\ref{dngrbtrue}). 
The probability of a given GRB  appearing at redshift $z$ is
\begin{equation}
P_z(z) = \frac{dN_{GRB}/dz}{\int_{0}^z(dN_{GRB}/dz)dz}.
\label{pz}
\end{equation}
The PDF was generated by $10^{5}$  random realizations based on  Eqs. (\ref{dngrbtrue}) and (\ref{pz}).  
Figure \ref{fig:probz} shows the  probability of finding a GRB at a given redshift, indicating that
a 50\%  probability of having  a GRB from a Pop~III star is obtained in the redshift range
$z \sim 7-11$ and a  95\%  probability in the range $z \sim 4-15$. Due to absorption from dust, IGM and DLA,  the GRBs available for observation by Gaia are restricted to the range $z \sim 3-7$. This results in approximately $10^4$ GRBs during the entire  Gaia nominal mission,  which is the value adopted in this work. 
 
\subsubsection{Half-opening angle PDF}

Using an empirical opening angle estimator, \citet{yonetoku2005} derived the opening angle PDF of GRBs. Their  PDF  can be fitted
by a power-lay ~ $\theta^{-2}$. Their results seem also  compatible with the universal structured jet model \citep{perna2003}.  The jet opening angle usually ranges between $1^{\circ}-10^{\circ}$ \citep{Frail2001,Cenko2009}. For simplicity, we assume a similar  power law in the range $\theta_{j}^{min} =0.01$ and $\theta_{j}^{max} = 0.2$ to determine the PDF  of $\theta_{j}$,

\begin{equation}
P_{\theta_{j}(\theta)}  \propto \theta^{-2}. 
\label{pj}
\end{equation}

\begin{figure}
\includegraphics[width=1\linewidth]{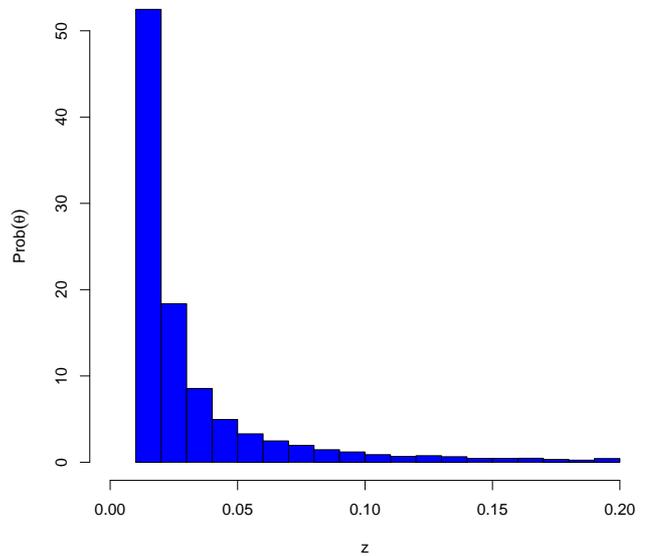}
\caption{Half-opening angle  jet PDF. Probability of a given GRB to have a particular  $\theta_{j}$. }
\label{fig:probtheta}
\end{figure}

Figure $\ref{fig:probtheta}$ shows the PDF  of $\theta_{j}$ generated by $10^{5}$ realizations based on Eq. (\ref{pj}). 
The realizations were performed within the range $\theta_{j} = 0.01-0.2$.  
The observational angle  $\theta_{obs}$ was randomly chosen between $0-\pi$.  From the relation $E_{\rm true} = \theta_{j}^2E_{\rm iso}/2$, we assume two fixed values for  $E_{true} = (2.5-5)\times 10^{50}$ ergs, which imposes the limits  $E_{iso } < 5\times 10^{54}-10^{55} ergs$  respectively.

\section{The Gaia mission}

\begin{figure}
\includegraphics[width=1\linewidth, angle = 0]{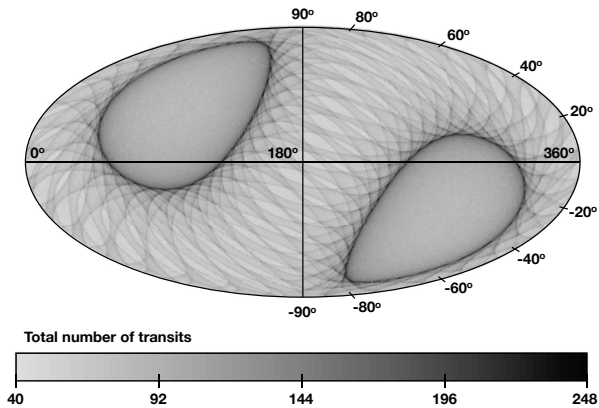}
\caption{Number of times each region of the sky (in galactic coordinates) will be observed by the Gaia satellite during the entire mission. }
\label{fig:GaiaNtransits}
\end{figure}

The Gaia satellite will perform observations of the entire sky, using a continuous scanning formed by the coupling of rotation and precession movements,   the scanning law. This law guarantees that each point in the sky will be observed several times during the mission, as it can be seen in Fig. \ref{fig:GaiaNtransits}.

Similar to what happens with CCD meridian circles, in the referential of the satellite's focal plane,  the sky continuously moves from one side to the other while the satellite spins. During this time, the CCD charges are synchronously transferred  to compensate for  the sky's apparent  motion and allow the integration.

This continuous observation strategy requires an equally continuous reading of the CCDs. Also, since Gaia's focal plane comprises 106 individual detectors\footnote{For a diagram of Gaia's focal plane, see,  e.g., \citet{Jordi:2010p7353}.}, it is not possible to transfer the entire content of the focal plane to the Earth due to bandwidth limits. So, a continuous analysis of the focal plane observations is also performed on-board, aimed at the detection of astronomical sources. When a source is detected, a rectangular ``window'' comprising a few arcseconds around the detected source is created (its exact size and pixel binning depend on the focal plane's CCD column). These windows are then transferred to the Earth.

For point sources, these observations will be unbiased and the data from all objects in the sky, bellow a certain limiting magnitude, will be sent to the ground. Certainly, among all those objects, not only galactic sources will be present, but also extragalactic ones. In particular, it is expected that point sources up to magnitude 20, in the Gaia passband G\footnote{This is a broad passband, which covers from 330-1000 nm. The nominal transmission curve can be found at \citet{Jordi:2010p7353}.}, will be ``windowed'' and transferred\footnote{After the mission (and during the mission for some problematic cases), it will be possible to reconstruct a deeper image around each detected source. In those reconstructed images, it will be possible to reach deeper magnitudes, albeit with some contamination from reconstruction artifacts.}.

As seen in Fig. \ref{fig:GRBafterglow}, some of the OA events are expected to remain above this limiting magnitude for a certain amount of time. The question that remains is whether their duration (at G$\leqslant$20) is enough for them to be observed at a reasonable rate. Only two quantities play an important role in  estimating the probability of Gaia observing  single event from a Pop III.2: the time whichen the OA remains brighter than G=20, $\Delta t$, and the coordinates $(l_{gal}, b_{gal})$ where the event takes place in the sky. Since those quantities are continuous distributions, it is necessary to analyze how the observation probability depends on them by building $P(\Delta t, l_{gal}, b_{gal})$. 
In the present work, we proceed as follows.

For a given coordinate in the sky, we start by computing the inverse Gaia scanning law  to derive a transit time list comprising the instants when Gaia's telescopes will be pointing at that coordinate. To be as realistic as possible, we adopt the Gaia Data Processing and Analysis Consortium's nominal implementation of it.
Then, we randomly select a point in time during the entire mission lifetime in order to place an event of a certain duration $\Delta t$. Using the transit time list,  we check if that event was observed, considering a time window of 4.4 seconds around each transit;  this is the time needed for the signal to cross the detection CCD and enter the confirmation CCD. If there is a superposition between the event duration and this time window, the event is considered detected. This procedure is then repeated until the estimation of the detection probability, which is derived by simply dividing the number of detected events by the total, does not vary more than 1\% between iterations. Finally, the whole procedure is repeated for each event duration $\Delta t$. As a consequence, we obtain an adequate time-sampling of the $P(\Delta t, l_{gal}, b_{gal})$ distribution.

For the determination of the number of OA events observed by Gaia on the entire sky, the coordinate dependency can be averaged out, allowing $P(\Delta t, l_{gal}, b_{gal}) \sim P(\Delta t) \pm \epsilon$. This is possible because the scanning law is mostly known, so  we can reasonably assume that the OA events take place randomly in the sphere.

The procedure described above was repeated for several positions on the sphere, and the mean and the standard deviation at each event duration were computed. To allow a good spatial sampling for the estimation of $P(\Delta t) \pm \epsilon$, we tessellate the celestial sphere at the hierarchical triangular mesh,  level 4 \citep{Kunszt:2001p8829}. This means that the simulations were performed at the center of 2048 triangles of approximately equal areas.

Finally,  to obtain the probabilities for the whole sky, an additional effect must be taken into account: the structure of our own Galaxy. Since the OAs are extragalactic events, the probability of observation at the galactic plane or bulge should be null or very small, due to  extinction and crowding. In this work, we conservatively assumed a null value for the probability of OAs being observed at such regions of the sky (defined here as $|b|\le15^\circ$ for $345^\circ\le l \le 15^\circ$ and $|b|\le5^\circ$ otherwise).

\begin{figure}
\includegraphics[width=1\linewidth, angle = 0]{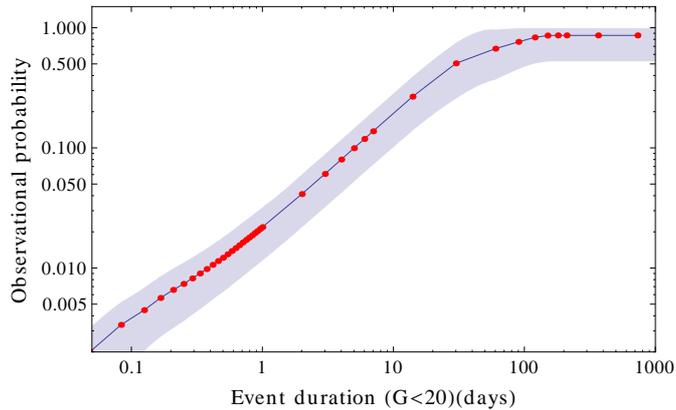}
\caption{Probability for a transient event with duration $\Delta t$ to be observed by Gaia. $\Delta t$ is the time the event stays brighter than the Gaia limiting magnitude during the 5 years nominal mission.}
\label{fig:sphere}
\end{figure}

The final results, representing the behavior of $P(\Delta t) \pm \epsilon$ can be seen in Fig. \ref{fig:sphere}.

\subsection{Analysis}

\begin{figure}
\includegraphics[width=1\columnwidth]{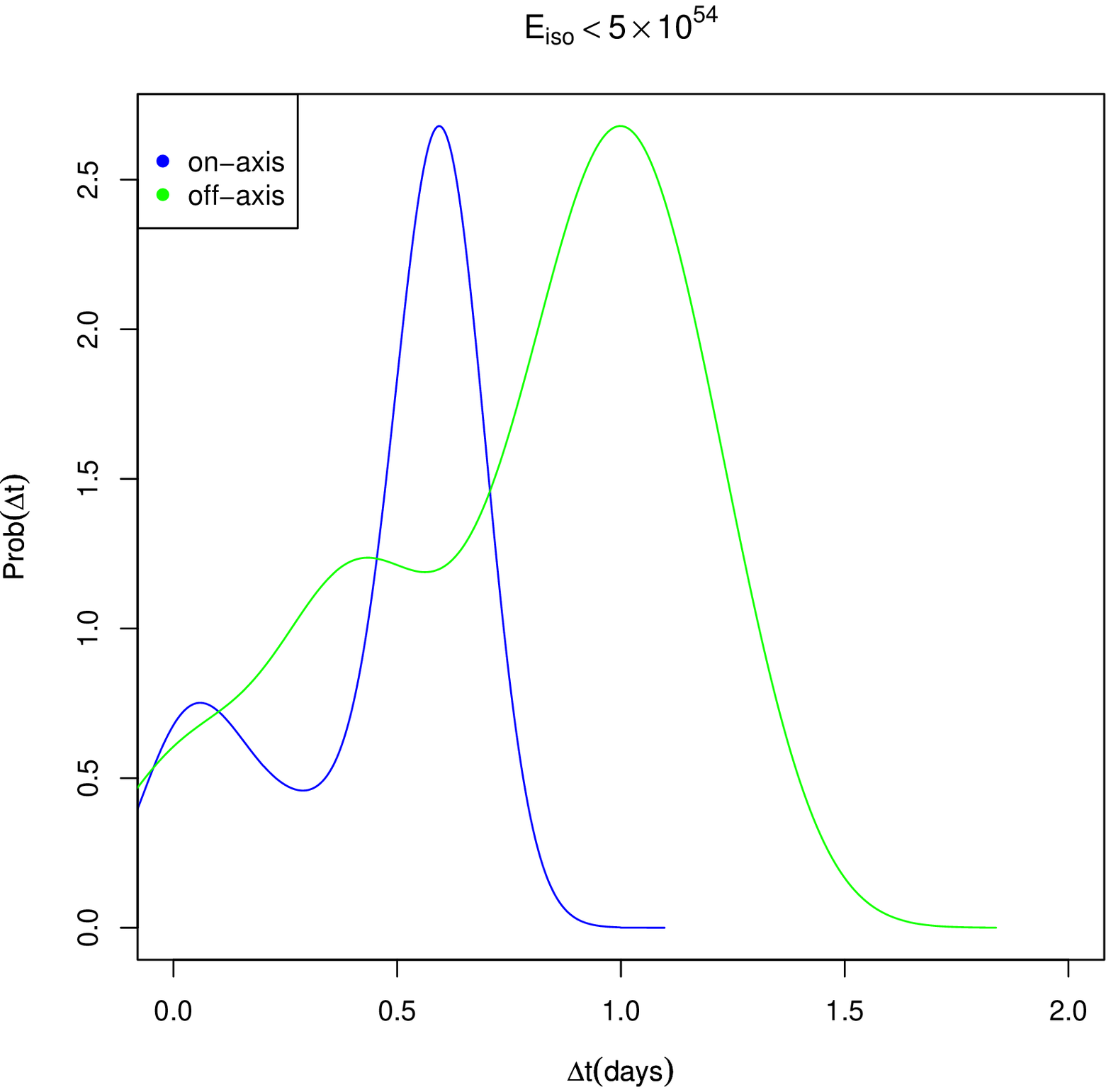}
\caption{PDF  of $\Delta t (days).$ Probability of an orphan afterglow with $E_{iso} \leqslant 5 \times10^{54}$  ergs  to appear above the Gaia flux limit for a given time interval. }
\label{fig:probdeltat54}
\end{figure}

\begin{figure}
\includegraphics[width=1\columnwidth]{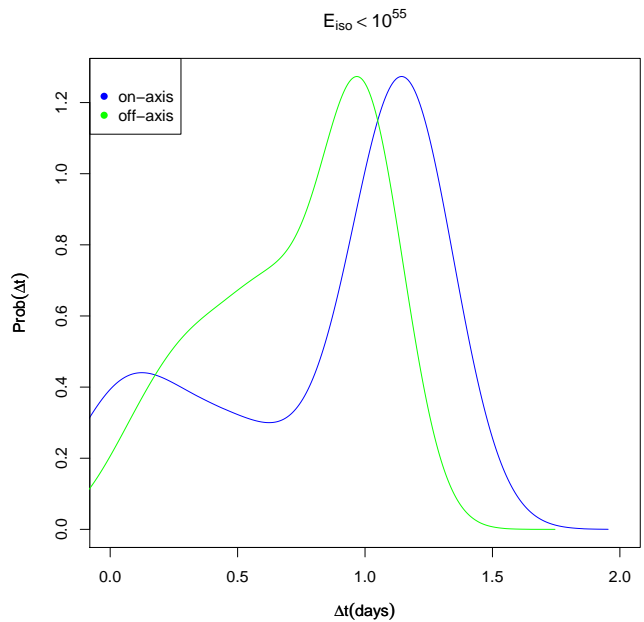}
\caption{PDF  of $\Delta t (days).$ Probability of an orphan afterglow with $E_{iso} \leqslant 10^{55}$  ergs  to appear above the Gaia flux limit for a given time interval. }
\label{fig:probdeltat55}
\end{figure}

In accordance with upper limit showed in Fig. \ref{fig:SFRII} and results from  \citet{rafael2011}, we expect between $\sim 10^2-5\times10^3$  events per year in all the  sky.   The uncertainties  come from our poor understanding about the efficiency  with which gas is  converted  into stars and  GRBs are triggered (two unknown factors for Pop III stars).  For a good statistics, we create a mock sample of $10^5$ events randomly generated by  the Monte Carlo method in order to infer the PDF  of an event to stay below $G = 20$ over $\Delta t (\rm days)$.  The average behavior for on-axis and off-axis afterglows as a function of $E_{iso}$ distribution  is shown in Figs. \ref{fig:probdeltat54}-\ref{fig:probdeltat55}.  Once we have $P (\Delta t)$, we can generate a sample with $10^4$ events several times and test against their probability of being observed by Gaia, given by Fig.  \ref{fig:sphere}.  Combining Figs. \ref{fig:sphere},  \ref{fig:probdeltat54}, and \ref{fig:probdeltat55}, we obtain the following results for the average number of events observed during the five  years of  the Gaia mission:

\begin{itemize}

\item $E_{iso} \leqslant 5\times 10^{54}$

 on-axis:1.34 $\pm$ 0.62, \\
 off-axis: 1.26 $\pm$ 0.53,

\item $E_{iso} \leqslant 10^{55}$

on-axis: 2.78 $\pm$ 1.41, \\ 
off-axis: 2.28 $\pm$ 0.88 

\end{itemize}

Despite the fact that the total number of on-axis is
always much lower than the number of off-axis, the
observed number depends on assumptions regarding the GRB luminosity functions. For lower energies, the decrease in flux due to the observation angle leads to a larger number  of light curves below the observational threshold. Thus, those on-axis have higher probability to be detected than those off-axis.


\section{Discussion}
\label{sec:conclusions}

Despite 
recent developments in theoretical studies
on the formation of  early generation of stars, there are no direct observations of Pop III stars yet.
Following the suggestion that massive Pop~III stars 
could trigger collapsar GRBs, we investigated the possibility to observe their OAs. 
We used previous results  from the literature to estimate the SFR 
for Pop III.2 stars, including all relevant feedback effects:  
photo-dissociation, reionization,  and metal enrichment. 
 
Since we expect a  larger number of OAs than on-axis GRBs, we estimated the possibility to observe such events during the five nominal operational years of the Gaia mission. The average number of events observed can be as high as  to 2.28 $\pm$ 0.88  off-axis afterglows and  2.78 $\pm$ 1.41  for on-axis ones. This implies that among the possible afterglows observed by Gaia \citep{Japelj2011},  a nonnegligible percentage ($\sim 10\%$)  might belong to Pop III stars. 
 
However, the detection of those events among the Gaia data will not be easy. Gaia will observe more than one billion objects all over the sky, and each object will be independently detected around eighty times during the mission, comprising a total of around $10^{12}$ astrometric, spectrophotometric,  and spectroscopic observations (after  detection, the observations are multiplexed in the focal plane). Consequently,   finding the OAs events among all that data can be a quite challenging task.


In  Gaia data processing,  a system called AlertPipe is being implemented to deal with alerts of transient events. It is foreseen to operate as follows: first candidate alerts are classified using Gaia data, then sources are cross-matched with available catalogues through Virtual Observatory or local copies and further classified,  and finally the alerts are stored on an alert server and released to the community as VOEvents \citep{HodgkinWyrzykowski2011, vanLeeuwen2011}. Algorithms are under analysis for dealing with GRBs (Wyrzykowski, 2012 priv. comm.), but no performance figures are available at the present moment.

Based on Gaia data, the duration of the OA can be roughly estimated from the flux variation between two subsequent observations: if the event is detected during the transit of the first telescope, it will be re-observed 106.5 minutes later when Gaia's second telescope re-observes the field. Moreover, the light curve will be sampled several times during the transit of each telescope, since at each column of Gaia's focal plane an independent magnitude measurement will be performed (measurements are spaced by 4.4s).

The light curve alone may not be enough to distinguish between GRB afterglows and other optical transient sources, as noted by \citet{Japelj2011}. However, as these events have power law like SEDs and no quiescent counterpart, this information should also be considered. Further analysis of Gaia's BP/RP low-dispersion spectrophotometry\footnote{BP/RP are Gaia spectrophotometers. BP works between 330-680nm with 4-32 nm/pixel and RP works between 640-1000nm with 7-15 nm/pixel.} are needed to distinguish between different transients with similar characteristics. To perform transient event classification, AlertPipe uses several algorithms,  including bayesian classifiers, template matching,  and self-organizing maps \citep{HodgkinWyrzykowski2010}.


A possible way  to search for such  objects within a large survey is to look for signatures of afterglows from Pop III stars. Two important characteristics of these objects are the total energy of Pop III GRBs, which  can be much higher than those of Pop I/II GRBs,  and the active duration time of their jet, which can be much longer than Pop I/II GRB jets due to the larger progenitor star.  Consequently, the detection of GRBs with very high $E_{iso}$ and very long duration could be indicative of such objects \citep{suwa2011,toma2011}. Thus,  they should appear as  quasi-steady point sources in  the radio survey observations. But the indication should be complemented with the constraint on the metal abundances in the surrounding medium with high-resolution IR and X-ray spectroscopy.  Since we do not have any observation of these objects,  we have to rely on theoretical models to compare the data. A way to look for such objects that is worth  future investigation is the use of   automatic light curve classifiers, which are  widely  implemented  for classifying  supernovae and transients in general \citep{johnson2006,kuznetsova2007,poznanski2007,rodney2009,falck2010,newling2011,richards2011,sako2011,ishida2012}. In principle, the theoretical model could work as a training set for the classifier, which would be then applied to surveys to identify 
possible candidates for further spectroscopical follow up. 


These OAs  event will be detected by the Gaia data processing pipeline just like any other transient. The timescales for raising the alerts are very dependent on specificities of the Gaia dataflow, but it is foreseen that, in the worst case, the data will be available for analysis by AlertPipe 24h after the observation. The alerts  will thus be raised no later than 48h after the Gaia observation \citep{WyrzykowskiHodgkin2011}. Nonetheless, it is not yet clear if AlertPipe by itself will be able to determine the nature of the transient as an OA.



Moreover, due to the design of the mission dataflow,  real-time identification will not be possible, but further identification of OAs using data from satellites/telescopes operating on other wavelengths may be possible, as VOEvents will be created by the Gaia data processing alert system. Also, OAs could be identified if they trigger X-ray detectors, such as Swift's BAT \citep{Barthelmy2005}, Fermi's LAT \citep{Atwood2009}, which is foreseen to operate until 2018, or future instruments, such as SVOM \citep{Schanne2010}. Finally, the same may also be observed by other large-scale optical surveys on Earth, e.g.,  LSST \citep{Ivezic2008} and Pan-STARRS \citep{Kaiser2002}, improving the sampling of the events' light curve and  providing information on other optical bands.


\section{Conclusion}

 It is important to emphasize that  our knowledge concerning first stars and their GRBs is still quite incomplete. 
Many of their properties (e.g.,  characteristic mass, SFR and efficiency to trigger GRBs) are still very uncertain, and more reliable information can only come once a detection is confirmed.  Recently, \citet{hosokawa2011}, performing state-of-the-art radiation-hydrodynamics simulations, 
showed that the typical mass of primordial stars could be $\sim 43 M_{\odot}$, i.e.,  less massive than originally expected by theoretical models. 
Their results,  though,  are affected by assumptions on the initial conditions. This confirms that we are far away from understanding all characteristics of these objects and any observation would be of paramount importance to improve theoretical models.

 In this work, we estimated the average number of OAs events originating from Pop III stars that the Gaia mission may observe to be up to   2.28 $\pm$ 0.88  off-axis afterglows and  2.78 $\pm$ 1.41   on-axis ones. In case such events are found among Gaia data, valuable physical properties associated with  the primordial stars of our Universe and their environment could  be constrained.

\begin{acknowledgements}
We are happy to thank K. Ioka  for the very fruitful suggestions and careful revision. We also thank Andrea Ferrara,  Andrei Mesinger, Andr\'e Moitinho,  Eduardo Amores, and  Laerte Sodr\'e,  for helpful discussions.  R.S.S. thanks the Brazilian agency FAPESP (2009/05176-4) and CNPq (200297/2010-4) for financial support. This work was supported by the World Premier International Research Center Initiative (WPI Initiative), MEXT, Japan. A.K.M. thanks the Portuguese agency FCT (SFRH/BPD/74697/2010) for financial support. E.E.O.I. thanks the Brazilian agencies CAPES (1313-10-0) and FAPESP (2011/09525-3) for financial support. We also thank the Brazilian INCT-A for providing computational resources through the Gina machine, as well as F. Mignard, X. Luri and the Gaia Data Processing and Analysis Consortium Coordination Unit 2 - Simulations for providing the scanning law implementation classes as well as the wrapper for the HTM sphere partitioning method. Finally, we thank the Scuola Normale Superiore of Pisa (SNS), Italy, and Centro Brasileiro de Pesquisas Fisicas (CBPF), Brazil, for their hospitality during part of the work on this paper. 
\end{acknowledgements}

\end{document}